# Polarization-tailored Fano interference in plasmonic crystals: A Mueller matrix model of anisotropic Fano resonance

S.K. Ray[+], S. Chandel[+], A. K. Singh, A. Kumar, A. Mandal. S. Misra

P. Mitra* and N. Ghosh*

*Department of Physical Sciences,*

*Indian Institute of Science Education and Research (IISER) Kolkata.*

*Mohanpur 741246, India*

*Corresponding authors: nghosh@iiserkol.ac.in, pmitra@iiserkol.ac.in

[+] These authors contributed equally to this work.

**ABSTRACT**

We present a simple yet elegant Mueller matrix approach for controlling the Fano interference effect and engineering the resulting asymmetric spectral line shape in anisotropic optical system. The approach is founded on a generalized model of anisotropic Fano resonance, which relates the spectral asymmetry to two physically meaningful and experimentally accessible parameters of interference, namely, the Fano phase shift and the relative amplitudes of the interfering modes. The differences in these parameters between orthogonal linear polarizations in an anisotropic system are exploited to desirably tune the Fano spectral asymmetry using pre- and post-selection of optimized polarization states. Experimental control on the Fano phase and the relative amplitude parameters and resulting tuning of spectral asymmetry is demonstrated in waveguided plasmonic crystals using Mueller matrix-based polarization analysis. The approach enabled tailoring of several exotic regimes of Fano resonance including the complete reversal of the spectral asymmetry. The demonstrated control and the ensuing large tunability of Fano resonance in anisotropic systems shows potential for Fano resonance-based applications involving control and manipulation of electromagnetic waves at the nano scale.

Fano resonance that exhibits a characteristic asymmetric spectral line shape, is a universal phenomenon observed throughout atomic, molecular [1, 2] optical [3-15], nuclear [16] and solid state systems [17]. The asymmetric spectral line shape arises due to the interference of a discrete state with a continuum of states and is modeled using the *Fano asymmetry parameter –q* [1,3,4]. Although, Fano resonance is regarded as a characteristic feature of interacting quantum systems, this can also been observed in classical optical phenomena because interference is ubiquitous in classical optics. Indeed, Fano-type spectral asymmetry has been observed in the scattered intensity from various optical systems, in plasmonic nanostructures, in electromagnetic metamaterials, in photonic crystals, in Mie scattering from dielectric objects etc. [3-15]. Fano resonances in such micro and nano optical systems have been the subject of intensive investigations due to their numerous potential applications like in sensing, switching, lasing, filters and robust color display, nonlinear and slow-light devices, invisibility cloaking, and so forth [2, 4, 5, 18-21]. Most of the aforementioned applications are known to critically depend upon the ability to control or modulate the asymmetry of the line shape by external means [4]. Thus, tuning the Fano resonance via some experimentally accessible parameters are highly desirable for realistic applications [4] as well as for fundamental studies [22].

In this letter, we present a Mueller matrix [23, 24] approach for controlling and tuning Fano resonance in anisotropic optical systems. A generalized model developed to interpret anisotropic Fano resonance suggests that the asymmetric spectral line shape can be desirably tuned by modulating two experimentally accessible parameters of interference, namely, the Fano phase shift and the relative amplitudes of the interfering modes. We demonstrate the validity of our model on waveguided plasmonic crystals where the resulting asymmetric spectral line shapes are studied using Mueller matrix based polarization analysis and controlled by pre and post selection of optimized polarization states of light.

We begin with a phenomenological model where a Fano-type spectral asymmetry in the scattered intensity naturally arises due to the interference of the complex Lorentzian field ($j^R(\omega)$) of a narrow resonance with a broad spectrum of relative field amplitude ($B$) assumed to be independent of frequency $\omega$ (ideal continuum):

$$E_s(\omega) \approx [j^R(\omega) + B] = \left[\frac{(q-i)}{(\epsilon+i)} + B\right] = \left[\frac{\sqrt{q^2+1}}{\sqrt{\epsilon^2+1}} e^{i\psi(\omega)} + B(\omega)\right] \quad (1)$$

Here, $\epsilon = \epsilon(\omega) = \frac{\omega-\omega_0}{(\gamma/2)}$, $\omega_0$ and $\gamma$ are the central frequency and the width of the narrow resonance, respectively. The phase difference between the interfering modes $\psi(\omega)$, comprises of two factors; the phase $\theta(\omega) = -tan^{-1}\left(\frac{1}{\epsilon}\right)$ associated with the narrow resonance and the Fano phase shift $\varphi_F$ [22], related to the asymmetry parameter $q$ as

$$\varphi_F = -tan^{-1}\left(\frac{1}{q}\right) \qquad (2)$$

The resulting expression for the scattered intensity becomes

$$I_s(\omega) = |E_s(\omega)|^2 \approx B^2 \times \left[\frac{(q^{eff}+\epsilon)^2}{(\epsilon^2+1)}\right] + \frac{(B-1)^2}{(\epsilon^2+1)} \qquad (3)$$

The first term represents the Fano-type asymmetric line shape with an effective asymmetry parameter $q^{eff} = q/B$. The second term corresponds to a Lorentzian background, widely reported in Fano resonance from diverse optical systems [6, 25]. It is worth emphasizing the crucial role of the parameters $(q - i)$ and $B$ in Eq. (1) that leads to the useful expression (Eq. 3) for Fano resonance in the scattered intensity with the emergence of an effective asymmetry parameter. This also enables intuitive interpretation of the essential spectral features by the two parameters, $\varphi_F$ and $B$ of the interfering scattered fields (see Supplementary information). The parameter $B$ controls the contrast of the Fano interference.

We now proceed to model anisotropic Fano resonance using Eq. (1). The scattered fields corresponding to the narrow resonance and the continuum can exhibit both phase and amplitude anisotropy, which can be modelled using the Jones matrix [23, 24]

$$E_s(\omega) = J(\omega)E_i; \quad J(\omega) = \begin{pmatrix} j_x^R + B_x & 0 \\ 0 & j_y^R + B_y \end{pmatrix}; \quad j_{x/y}^R = \frac{q_{x/y}-i}{\epsilon_{x/y}+i} \qquad (4)$$

Here, $x/y$ denotes orthogonal linear polarizations. The phase and amplitude anisotropy of the narrow resonance is included as $q_x \neq q_y$ (or differences in the Fano phase between orthogonal linear polarizations, $\varphi_{F,x} - \varphi_{F,y}$) and $\epsilon_x \neq \epsilon_y$ (which is primarily due to differences in the resonance frequencies $\omega_{0,x} \neq \omega_{0,y}$). Similarly, $B_x \neq B_y$ results in the anisotropy of the continuum. The corresponding 4×4 scattering Mueller matrix (M) describing the Stokes vector (S) transformation is a diattenuating retarder [23, 26, 27] having magnitudes of linear diattenuation $d$ (differential attenuation) and linear retardance $\delta$ (phase difference between orthogonal linear polarizations) given by

$$d = \frac{|J_{11}|^2 - |J_{22}|^2}{|J_{11}|^2 + |J_{22}|^2} = \frac{M_{12}}{M_{11}}; \quad \delta = \arg(J_{11}) - \arg(J_{22}) = tan^{-1}\frac{M_{34}}{M_{44}} \qquad (5)$$

The amplitude anisotropy (encoded in $d$) and the phase anisotropy (encoded in $\delta$) can be directly determined from experimental Mueller matrices, as described subsequently.

Equations (1, 3, and 4) reveal that the polarization state offers a convenient handle to control the $\varphi_F$ and $B$ parameters and the resulting $q^{eff}$ parameter in anisotropic system. This can be achieved by pre- and post-selection of polarization states and the corresponding expression for the scattered intensity becomes

$$I_s(\omega) = \left|\mathbf{E}_\beta^+ \, \mathbf{J} \, \mathbf{E}_\alpha\right|^2 = \tfrac{1}{2} \mathbf{S}_\beta^T \, M \, \mathbf{S}_\alpha \qquad (6)$$

Here, $\mathbf{E}_\alpha = [\cos\alpha \quad \sin\alpha \, e^{i\Phi_\alpha}]^T$, $\mathbf{E}_\beta = [\cos\beta \quad \sin\beta \, e^{i\Phi_\beta}]^T$ are Jones vectors for general elliptical polarizations and $\mathbf{S}_\alpha$, $\mathbf{S}_\beta$ are their corresponding Stokes vectors [23, 24]. Two types of anisotropic systems are of potential importance: *Type- (a)*, where the narrow resonance exhibits both phase and amplitude anisotropy ($q_x \neq q_y, \epsilon_x \neq \epsilon_y$) [10-12], and *Type- (b)*, where the narrow resonance is perfectly diattenuating - $j_x^R \neq 0$ and $j_y^R = 0$ [8-9]. Eq. (6) can be approximated in the form of Eq. (3) with effective asymmetry parameter ($q^{eff}$) and effective relative amplitude parameter $B^{eff}$. In type (a), $q^{eff}$ can be controlled by directly modifying $\varphi_F$ (or $q = -\cot\varphi_F$) using pre and post-selections in linear polarization ($\Phi_\alpha = \Phi_\beta = 0$). In the limit $(\omega_{0,x} - \omega_{0,y}) \ll \gamma$, the corresponding parameters can be approximated as

$$q^{eff} = \tfrac{q}{B}, \; q \approx \left[\frac{(q_x \times \cos\alpha \cos\beta + q_y \times \sin\alpha \sin\beta)}{\cos(\alpha-\beta)}\right] \qquad (7)$$

In type (b), $q^{eff}$ can be controlled by changing $B^{eff}$ with

$$q^{eff} = \tfrac{q}{B^{eff}}, \; B^{eff} = \left(1 + \frac{\sin\alpha \sin\beta}{\cos\alpha \cos\beta}\right)B \qquad (8)$$

For simplicity, we have assumed the continuum mode to be isotropic ($B_x = B_y$).

While the above analysis is valid for non-depolarizing interactions, this can be implemented in depolarizing situations also. The depolarizing component of the scattered light can be filtered out using decomposition of Mueller matrix into basis matrices of a depolarizer ($M_{Depol}$) and a non-depolarizing diattenuating retarder ($M_{Pol}$) [23, 26 - 28], and the latter can be subjected to analysis using Eq. 6 (see Supplementary information). In what follows, we experimentally demonstrate our approach in waveguided plasmonic crystals [8-12].

The waveguided plasmonic crystal samples consisted of two dimensional periodic array of gold (Au) nano-disks (or nano-ellipse) on top of ~ 190 nm thick indium tin oxide (ITO) waveguiding layer coated on quartz substrate. We used electron beam lithography and metal deposition by thermal evaporation technique to fabricate these nanostructures (see Supplementary information). The dimension of the fabricated Au circular disk array was (diameter D = 160 nm, height = 30 nm, centre to centre distance L = 480 nm) and the corresponding dimension of the Au ellipse array was ($D_x$ = 134 nm, $D_y$ =95nm, height = 30 nm, L = 480 nm).

The elastic scattering Mueller matrices $M$ from the samples were recorded using a home-built spectroscopic Mueller matrix system integrated with an inverted microscope operating in the dark-field mode (see Supplementary Figure S1) [26]. It employs broadband white light excitation and subsequent recording of sixteen polarization resolved scattering spectra (wavelength λ = 400 – 725 nm, 1 nm resolution) by sequential generation and analysis of four optimized elliptical polarization states (see Supplementary information) [26].

Typical SEM and dark field images of Au circular disk array are shown in Fig. 1a and 1b. Such waveguided plasmonic crystals exhibit strong coupling between the localized plasmons (of metallic nanostructures) and waveguiding modes (in ITO dielectric layer) [8 – 12, 29, 30]. In the presence of the periodic metal nanostructures, the bound guided modes will couple to the photon continua and become leaky (quasiguided) [29, 30]. The interference of the scattered fields of these quasiguided modes (acting as the narrow resonance peaking at $E_0 = \hbar\omega \sim 1.777\ eV$, $\lambda_0 \sim 698\ nm$) and the dipolar plasmon resonance of the Au disk array (acting as continuum) leads to Fano resonance in the scattering spectra (peak at $E_m \sim 1.896\ eV$, $\lambda_m \sim 654 nm$). This is evident from the observed spectral asymmetry in the polarization-blind scattering spectra ($M_{11}$ element) from Au circular disk array (Fig. 1c). Fitting with Eq. (3) yields the effective asymmetry parameter $q^{eff} \sim +0.90$ and the Fano phase $\varphi_F \sim -63.2°$ (determined using Eq. 2). The corresponding parameters in a similar Au ellipse array (Fig. 1c) were $q^{eff} \sim +1.562 \rightarrow \varphi_F \sim -47.5°$. For the experimental demonstration of the proposed approach (based on Eq. 6-8), the continuum mode is desired to be isotropic. The Au circular disk array instead of the ellipse array is therefore chosen for subsequent Mueller matrix analysis.

The anisotropy of Fano resonance is manifested as non-zero off-diagonal elements in the Mueller matrix $M$ of the Au circular disk array (Fig. 2a). Additionally, $M$ contains depolarization contributions (due to averaging effects in high NA microscopic geometry and reflected in the diagonal elements, whose magnitudes are less than unity), which are efficiently filtered out (via $M_{Depol}$) using polar decomposition [28] (see Supplementary Fig. S2). The resulting non-depolarizing $M_{Pol}$ matrix is used to glean the phase and the amplitude anisotropy effects, signatures of which are characteristically encoded in the $M_{34}$ / $M_{43}$, $M_{24}$/ $M_{42}$ elements and $M_{12}$ / $M_{21}$, $M_{13}$/ $M_{31}$ elements, respectively. These are subsequently quantified through linear retardance $\delta$ and diattenuation $d$ parameters (Fig. 2b) (using Eq. S5 of Supplementary information). The scattered fields corresponding to the $TM$ / $TE$ quasi-guided modes (excited by $x/y$ linear polarizations, respectively, illustrated in Fig. 1d) exhibit both phase difference ($\varphi_{F,x} \neq \varphi_{F,y}$) and amplitude difference (due to both $q_x \neq q_y$ and $E_{0,x} \neq E_{0,y}$) between orthogonal linear polarizations. This differential polarization response of the quasiguided modes is the primary source of the Fano resonance anisotropy. This is evident from the corresponding rapid variation of $\delta$ and $d$ across the narrow resonance peak of the quasiguided modes ($E \sim 1.8\ eV\ or\ \lambda \sim 690\ nm$). Moreover, as envisaged from Eq. (4) and (5), the rapidly varying components ($\delta_{Fano}$ and $d_{Fano}$ in Fig. 2b) around the narrow spectral range of the asymmetry are only pertinent to anisotropic Fano resonance (the $\delta$ parameter additionally possesses a frequency-independent background $\sim 0.8$ *rad*).

We now turn to inspect whether the spectral asymmetry parameters ($q_x, q_y$) are indeed linked to the measured anisotropy parameters, as predicted by Eq. 4 and 5. The $M_{Pol}$ matrix is thus utilized further to determine the *q*-parameters for the two orthogonal linear polarizations ($TM - x$ and $TE - y\ polarization$) (Fig. 3a and 3b). Differences in the *q*-parameters and the Fano phases between *x*-polarization ($q_x^{eff} = 0.931, q_x = 0.542 \rightarrow \varphi_{F,x} \sim -61.5°$) and *y*-polarizations ($q_y^{eff} = 0.861, q_y = 0.471 \rightarrow \varphi_{F,y} \sim -64.8°$) confirm anisotropic nature of the spectral asymmetry, albeit with relatively weaker magnitude (small differences in $\varphi_{F,x} - \varphi_{F,y}$, and $E_{0,x} - E_{0,y}$). The estimated Fano phase and amplitude parameters ($q_x/q_y, B_x/B_y$) were subsequently used to predict (using Eq. 4 and 5) the spectral variations of the $\delta$ and $d$ parameters (Fig. 3c and 3d), which show good agreement with the corresponding variations obtained directly from the experimental Mueller matrix ($\delta_{Fano}$ and $d_{Fano}$ in Fig. 2b). This establishes self-consistency and demonstrates that the Mueller matrix approach enables one to check the accuracy of the Fano asymmetry parameters ($q_x/q_y, q_x^{eff}/q_y^{eff}$), which are determined by fitting the spectral variation of the intensity to Eq. 3.

The above results demonstrate that the experimentally observed spectral asymmetry ($q^{eff}$) can be mapped to two parameters of interference, $\varphi_F\ and\ B$. The Fano resonance anisotropy originating from the differences in these parameters for orthogonal polarizations, leaves its unique signature as rapidly varying spectral retardance ($\delta$) and diattenuation ($d$) effects in the Mueller matrix. The $\delta\ and\ d$ parameters thus hold promise as novel experimental metric for probing and analyzing anisotropic Fano resonance. The general recipe provided by our model (Eq. 6-8) may now be explored for tuning of Fano resonance in anisotropic systems.

Tuning of the Fano asymmetry parameter ($q^{eff}$) by pre and post selection of optimized polarization states is illustrated in Fig. 4a. Here, the pre-selected state is optimized to be elliptical ($\alpha = +45^0, \Phi_\alpha - \Phi_\beta \sim 0.8 \; rad$) so that the additional background anisotropy (retardance $\delta \sim 0.8 \; rad$, see Fig. 2b) of the sample is compensated. The post-selections are at varying linear polarization angles $\beta$. This is equivalent to the *Type-(a)* anisotropic system with pre and post-selection in linear polarization basis. In conformity with the corresponding predictions (Eq. 7), the results demonstrate tuning of $q^{eff}$ by modifying the Fano phase $\varphi_F$ (or $q$), albeit for a limited range permitted by the moderate level of the anisotropy. Note that the $B^{eff}$ parameter also varies here due to $B_x \neq B_y$ (as noted in Fig. 3a and 3b) (see Fig. S3 in Supplementary). Interestingly, even for this weakly anisotropic system, the spectral asymmetry is reversed ($q^{eff}: positive \rightarrow negative$) with a choice of $\beta \sim 130^o$ ($\varphi_F$ is reversed: $-61^o$ for $\beta = 0^o \rightarrow +88^o$ for $\beta = 130^o$). This is of particular practical interest because potentially a Fano spectral dip (energy / wavelength $E_F / \lambda_F$ corresponding to the intensity minima) can be turned to a spectral peak or vice versa, enabling a large tunability of $E_F / \lambda_F$. In this example, tunability $\sim$ 60 nm is achieved ($\lambda_F \sim 720 \; nm$ for $\beta = 0^o$ to $\lambda_F \sim 660 \; nm$ for $\beta = 130^o$). These features can be seen more prominently in the Fano resonant part of the fitted spectral intensity (first term in Eq. 3, Supplementary Fig. S4). Note that the anisotropy effects of the plasmonic crystals can be enhanced by changing the periodicity of the disk array [29]. As demonstrated in Fig. 4b, a much more dramatic control on $\varphi_F$ and the $q^{eff}$ parameters can be obtained in such strongly anisotropic system having larger differences in $\varphi_{F,x} - \varphi_{F,y}$ and $E_{0,x} - E_{0,y}$. The theoretical predictions (using $\alpha = +45^0$, varying $\beta$ in Eq. 4 and 6) reveal that several exotic regimes of Fano resonance can be achieved – (1) starting from a moderate spectral asymmetry, high degree of asymmetry can be obtained ($q^{eff} \sim +0.83, \varphi_F = -63^o$ for $\beta = 90^o$), (2) symmetric Lorentzian dip in the scattering spectra can be tailored ($q^{eff} \rightarrow 0, \varphi_F \rightarrow 90^o$ for $\beta = 110^o$), (3) the symmetric dip can be reversed to a Lorentzian peak ($q^{eff} \rightarrow \infty, \varphi_F \sim 0^o$ for $\beta = 135^o$), and importantly, (4) the spectral asymmetry can be completely reversed ($q^{eff} \sim -1.21, \varphi_F = +26^o$ for $\beta = 120^o$). This enables a large ($\sim$ 80 nm) tunability in the Fano spectral dip ($\lambda_F \sim 690 \; nm$ for $\beta = 90^o$ to $\lambda_F \sim 610 \; nm$ for $\beta = 120^o$). A similar type of control on Fano resonance can also be obtained in *Type-(b)* anisotropic system, where $q^{eff}$ can be controlled by changing $B^{eff}$ (see Supplementary Fig. S5). These are illustrative examples and many other interesting possibilities emerge, wherein the anisotropy and the spectral parameters of the narrow resonance can be appropriately designed to enable much larger tunability of Fano spectral dip. The relevant parameters ($q_x/q_y$, $B_x/B_y$) can be extracted from the Mueller matrix and subsequently used in combination with Eq. (6-8) to optimize the polarization states for tailoring the desirable spectral asymmetry ($q^{eff}$).

In conclusion, we have presented a Mueller matrix method for controlling the Fano interference effect and engineering the asymmetric spectral line shape in anisotropic system. The method is founded on a model of anisotropic Fano resonance that provides a general recipe for desirably tuning the spectral asymmetry by modulating two parameters of interference (Fano phase and relative amplitude) using pre and post selection of optimized polarization states. The principle is demonstrated on waveguided plasmonic crystals exhibiting moderate level of anisotropy, and a much more dramatic control is envisaged in strongly anisotropic systems. The large tunability in the Fano spectral dip in appropriately designed strongly anisotropic system may potentially enhance active Fano resonance-based applications [4, 19, 31-33], e.g., by enabling polarization-optimized multi-sensing platform [31, 32], developing polarization-controlled novel Fano switching devices [4, 19], extending applications in filtering and robust color display [33], and so forth.

**References**
1. U.Fano, *Phys. Rev.* **124,** 1866–1878 (1961).
2. F. Shafiei, C. Wu, Y. Wu, A. B. Khanikaev, P. Putzke, A. Singh, X. Li, and G. Shvets, *Nature Photon.* **7,** 367-372 (2013).
3. A. E. Miroshnichenko, S. Flach, and Y. S. Kivshar, *Rev. Mod. Phys.* **82,** 2257 (2010).
4. B. Luk'yanchuk, N. I. Zheludev, S.A.Maier, N. J. Halas, P. Nordlander, H. Giessen and C.T Chong, *Nature Mater.* **9,** 707-715 (2010).
5. C.Wu, A. B. Khanikaev and G.Shvets, *Phys.Rev. Lett.* **106,** 107403 (2011).
6. B.Gallinet, &O. J. F. Martin, *Phys. Rev. B* **83,** 235427 (2011).
7. J. A. Fan, C. Wu, K. Bao, J. Bao, R. Bardhan, N. J. Halas, V. N. Manoharan, P. Nordlander, G. Shvets, and F. Capasso, *Science.* **328,** 1135-1138 (2010).
8. M. R. Shcherbakov, P. P. Vabishchevich, V. V. Komarova, T. V. Dolgova, V. I. Panov, V. V. Moshchalkov, and A. A. Fedyanin, *Phys. Rev. Lett.* **108,** 253903 (2012).

9. A.Christ, S. G. Tikhodeev, N. A. Gippius, J.Kuhl, H.Giessen, *Phys. Rev. Lett.* **91,** 183901 (2003).
10. Y.Zhu, X. Hu, Y.Huang, H.Yang, and Q.Gong, *Adv. Optical Mater.* **1,** 61-67 (2013).
11. G.Gantzounis, N.Stefanou, and N. Papanikolaou, *Phys. Rev. B.* **77,** 035101 (2008).
12. M. Lisunova, J. Norman, P. Blake, G. T Forcherio, D. F. D. Jarnette and D. K. Roper, *J. Phys. D* **46,** 48 5103 (2013).
13. *Y. Sonnefraud, N. Verellen, H. Sobhani, G. A. E. Vandenbosch, V. V. Moshchalkov, P. V. Dorpe, P. Nordlander and S. A. Maier, ACS Nano.* **4,** 1664-1670 (2010).
14. X. Fan, W.Zheng, and D.J. Singh, *Light, Science and Application.* **3,** e179 (2014).
15. A.C.Johnson, C.M. Marcus, M.P.Hanson, and A.C.Gossard, *Phys. Rev. Lett.* **93,** 106803 (2004).
16. A. R. Schmidt, M. H. Hamidian, P. Wahl, F. Meier, A. V. Balatsky, J. D. Garrett, T. J. Williams, G. M. Luke and J. C. Davis, *Nature.* **465,** 570-576 (2010).
17. P.Fan, Z.Yu, S.Fan, &M. L. Brongersma, *Nature Mater.* **13,** 471- 475 (2014).
18. J. N. Anker, W. P. Hall, O.Lyandres, N. C. Shah, J. Zhao and R. P. Van Duyne, *Nature. Mater.* **7,** 442–453 (2008).
19. K. Nozaki, A. Shinya, S. Matsuo, T. Sato, E. Kuramochi and Masaya Notomi, Optics Express. 21(10), 11877-11888 (2013).
20. N. I. Zheludev, S. L. Prosvirnin, N.Papasimakis, andV. A. Fedotov, *Nat. Photon.* **2,** 351–354 (2008).
21. B. Zhang, *Light: Science & Applications.* **1,** e32 (2012).
22. C. Ott, A. Kaldun, P. Raith, K. Meyer, M. Laux, J. Evers, C. H. Keitel, C. H. Greene, and T. Pfeifer, *Science.* **340,** 716-720 (2013).
23. S.D. Gupta, N.Ghosh, &A.Banerjee, *Wave Optics: Basic Concepts and Contemporary Trends (CRC Press, 2015).*
24. D. H. Goldstein, *Polarized Light, Third Edition (CRC Press, 2010).*
25. B.Gallinet, andO. J. F. Martin, *ACS. Nano .***5,** 11, 8999-9009 (2011).
26. S. Chandel, J. Soni, S. K. Ray, A. Das, A. Ghosh, S. Raj, and N. Ghosh, *Sci. Rep.* **6,** 26466 (2016).
27. J. Soni, S. Ghosh, S. Mansha, A. Kumar, S. Dutta Gupta, A. Banerjee, and N. Ghosh, *Opt. Lett.***38 (10),** 1748-1750 (2013).
28. S. Y. Lu, and R. A.Chipman, *J. Opt. Soc. Am. A***13**, 1106 - 1113 (1996).
29. A. Christ,T. Zentgraf, J. Kuhl, S. G. Tikhodeev, N. A. Gippius, H. Giessen, Phys. Rev. B **70,** 125113 (2004)

30. S. G. Tikhodeev, A. L. Yablonskii, E. A. Muljarov, N. A. Gippius, Teruya Ishihara, Phys. Rev. B **66,** 045102 (2002).
31. C. Wu, A. B. Khanikaev, R. Adato, N. Arju, A. A. Yanik, H. Altug & G. Shvets, Nat. Matter.**11,** 69-75(2012).
32. A. Lovera, B. Gallinet, P. Nordlander and O. J.F. Martin, ACS Nano. **3,** 643–652 (2009).
33. Z. Zhang, A. W. Bargioni, S.W.Wu, S.Dhuey, S.Cabrini and P. J. Schuck, Nano Lett.9 (12), 4505–4509 (2009).

**Figure Captions**

**Figure 1:** Typical **(a)** SEM and **(b)** dark field images of Au circular disk array. **(c)** The scattering spectra from Au circular disk array (green dashed line) and ellipse array (red solid line) and theoretical fits of the spectra (for $E = \hbar\omega = 1.710 - 2.066$ eV) with Eq. (3) (dotted lines). The values of the parameters ($q^{eff}$, $B$) are noted. The peaks of the narrow Lorentzian resonance and the resulting Fano resonance were at ($E_0 = 1.777\ eV$, $\lambda_0 = 698$ nm; $E_m = 1.896$ eV, $\lambda_m = 654$nm) and ($E_0 = 1.784\ eV$, $\lambda_0 = 696$ nm; $E_m = 1.850$ eV, $\lambda_m = 671$ nm) for the circular disk and the ellipse array, respectively. **(d)** Illustration of the origin of *anisotropic* Fano resonance in waveguided plasmonic crystals.

**Figure 2: (a)** The experimental Mueller matrix $M$ ($E = 1.710 - 2.254$ eV shown here) of the Au circular disk array. The elements are normalized by the $M_{11}$ element. **(b)** The spectral variation of the derived linear diattenuation $d$ (blue solid line, left axis) linear retardance $\delta$ (red dashed line, right axis). The magnitudes of the rapidly varying components $d_{Fano}$ and $\delta_{Fano}$ across the peak of the narrow resonance are noted.

**Figure 3:** The spectral variation of scattered intensities for **(a)** $TM - x\ polarization$ and **(b)** $TE - y\ polarization$. These are obtained by pre and post selection (using Eq. 6) of corresponding polarization states (Stokes vector: $[1\ 1\ 0\ 0]^T$ for $x$ and $[1\ -1\ 0\ 0]^T$ for $y$) on the $M_{Pol}$ matrix. The values for ($q_x^{eff}$, $q_x, B_x$)/ ($q_y^{eff}$, $q_y, B_y$) parameters estimated by fitting to Eq. 3 (dotted lines) are noted. The resonance parameters were ($E_{0,x} = 1.773\ eV$, $\lambda_{0,x} = 699$ nm; $E_{m,x} = 1.890$ eV, $\lambda_{m,x} = 656$ nm) for $x$ and ($E_{0,y} = 1.779\ eV$, $\lambda_{0,y} = 697$ nm; $E_{m,y}=1.904$ eV, $\lambda_{m,y} = 651$ nm) *for y polarization.* The theoretical predictions (using Eq. 4 and 5) of spectral variation of **(c)** linear diattenuation $d_{Fano}$ and **(d)** linear retardance $\delta_{Fano}$.

**Figure 4:** Tuning of Fano spectral asymmetry in **(a, b)** moderately anisotropic experimental waveguided plasmonic crystal sample, and **(c, d)** theoretical predictions for a strongly anisotropic system. The pre-selected state is elliptical ($\alpha = +45^0, \Phi_\alpha - \Phi_\beta \sim 0.8\ rad$, $\mathbf{S}_\alpha = [1\ 0\ 0.697\ -0.717]^T$) in **(a, b)** and linear ($\alpha = +45^0, \Phi_\alpha - \Phi_\beta = 0$, $\mathbf{S}_\alpha = [1\ 0\ 1\ 0]^T$) in **(c, d)**. Post-selections are at different linear polarization angles β ($\mathbf{S}_\beta = [1\ \cos2\beta\ \sin2\beta\ 0]^T$). The spectral variation of the scattered intensities are shown in (a) and (c) and the changes in the line shapes are highlighted in (b) and (d). The fitted (using Eq. 3) parameters ($q^{eff}$, $\varphi_F$) are noted. In (a, b), the pre and post-selections are done on experimental $M_{Pol}$ matrix. In (c, d), these are simulated using Eq. (4) and (6) with ($q_x = 1.5, q_y = 0.5, B_x = B_y = B = 0.6$) and ($E_{0,x} = 1.893\ eV$, $\lambda_{0,x} = 655$ nm; $E_{0,y} = 1.882\ eV$, $\lambda_{0,y} = 659$ nm).

**Figures**

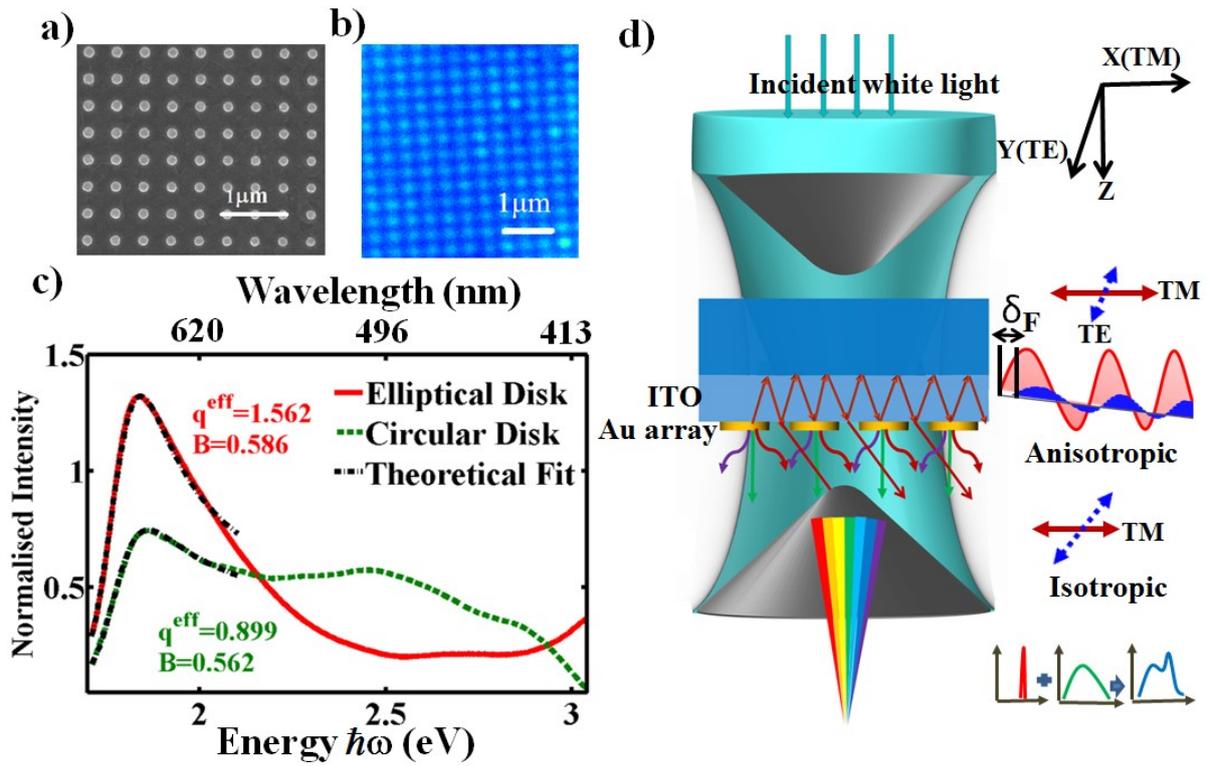

**Figure 1:** Typical **(a)** SEM and **(b)** dark field images of Au circular disk array. **(c)** The scattering spectra from Au circular disk array (green dashed line) and ellipse array (red solid line) and theoretical fits of the spectra (for $E = \hbar\omega = 1.710 - 2.066$ eV) with Eq. (3) (dotted lines). The values of the parameters ($q^{eff}$, $B$) are noted. The peaks of the narrow Lorentzian resonance and the resulting Fano resonance were at ($E_0 = 1.777\ eV$, $\lambda_0 = 698$ nm; $E_m = 1.896$ eV, $\lambda_m = 654$ nm) and ($E_0 = 1.784\ eV$, $\lambda_0 = 696$ nm; $E_m = 1.850$ eV, $\lambda_m = 671$ nm) for the circular disk and the ellipse array, respectively. **(d)** Illustration of the origin of *anisotropic* Fano resonance in waveguided plasmonic crystals.

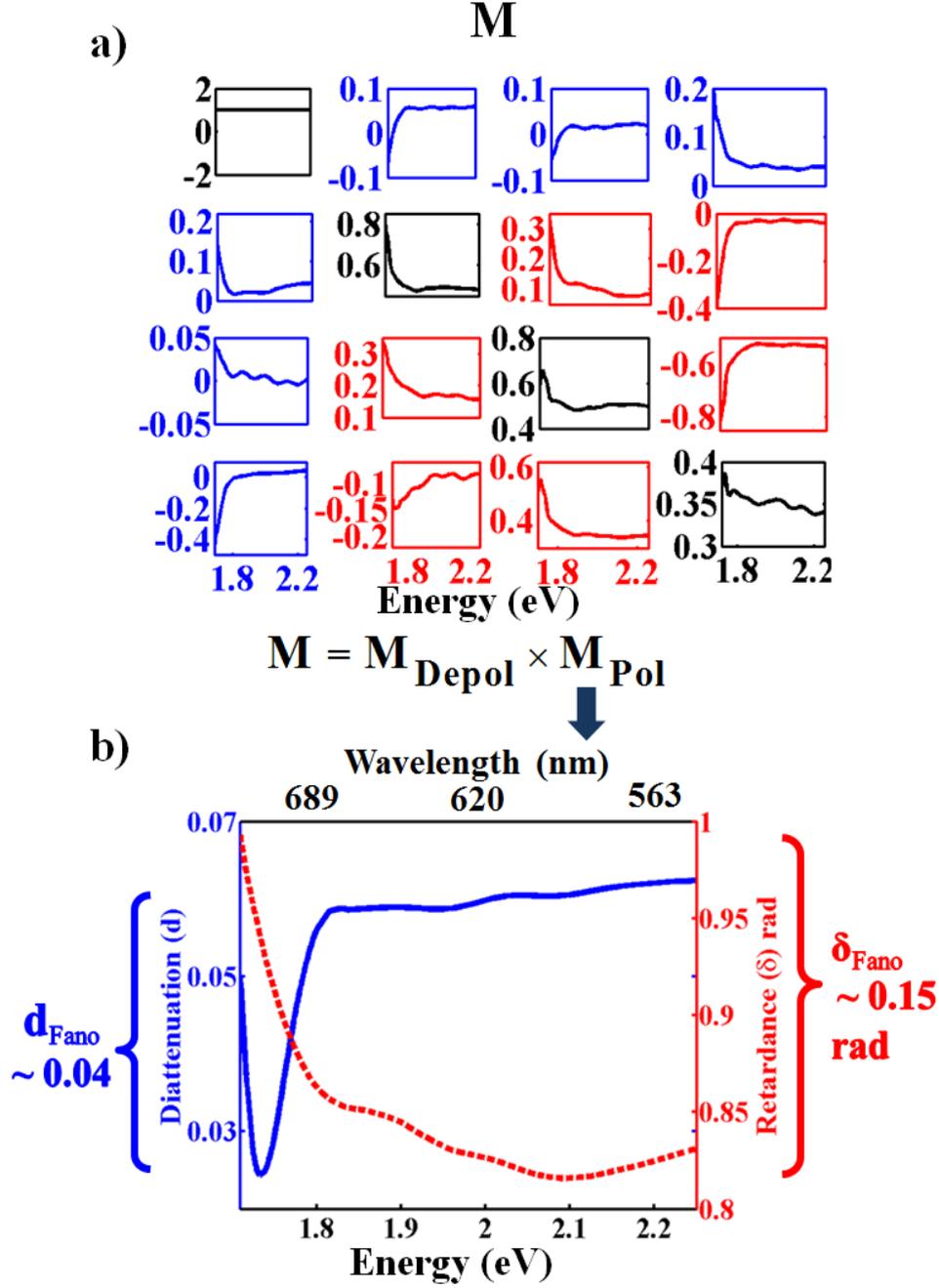

**Figure 2:** (a) The experimental Mueller matrix $M$ ($E = 1.710 – 2.254$ eV shown here) of the Au circular disk array. The elements are normalized by the $M_{11}$ element. (b) The spectral variation of the derived linear diattenuation $d$ (blue solid line, left axis) linear retardance $\delta$ (red dashed line, right axis). The magnitudes of the rapidly varying components $d_{Fano}$ and $\delta_{Fano}$ across the peak of the narrow resonance are noted.

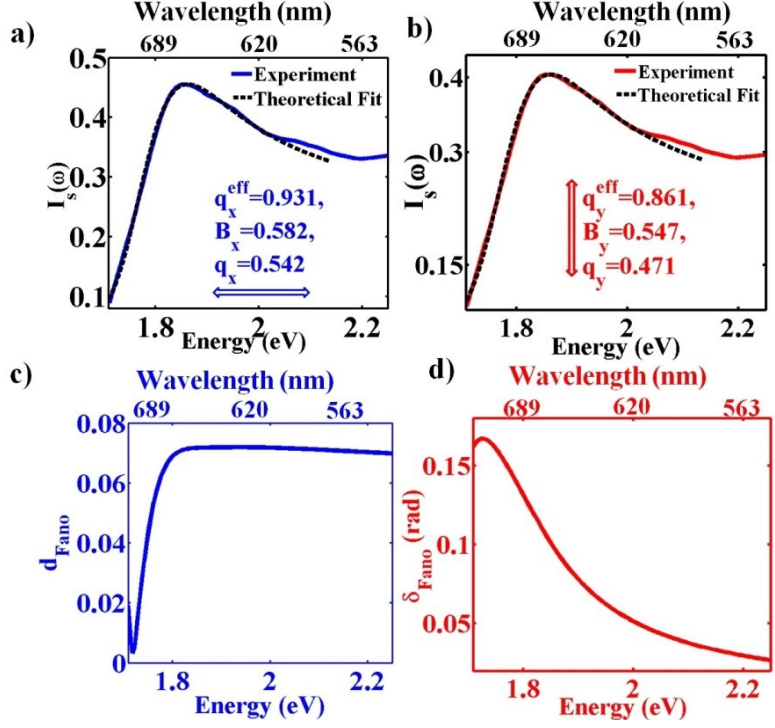

**Figure 3:** The spectral variation of scattered intensities for **(a)** $TM - x\ polarization$ and **(b)** $TE - y\ polarization$. These are obtained by pre and post selection (using Eq. 6) of corresponding polarization states (Stokes vector: $[1\ 1\ 0\ 0]^T$ for $x$ and $[1\ -1\ 0\ 0]^T$ for $y$) on the $M_{Pol}$ matrix. The values for $(q_x^{eff}, q_x, B_x)/(q_y^{eff}, q_y, B_y)$ parameters estimated by fitting to Eq. 3 (dotted lines) are noted. The resonance parameters were ($E_{0,x} = 1.773\ eV$, $\lambda_{0,x} = 699$ nm; $E_{m,x} = 1.890$ eV, $\lambda_{m,x} = 656$ nm) for $x$ and ($E_{0,y} = 1.779\ eV$, $\lambda_{0,y} = 697$ nm; $E_{m,y}=1.904$ eV, $\lambda_{m,y} = 651$ nm) *for y polarization.* The theoretical predictions (using Eq. 4 and 5) of spectral variation of **(c)** linear diattenuation $d_{Fano}$ and **(d)** linear retardance $\delta_{Fano}$.

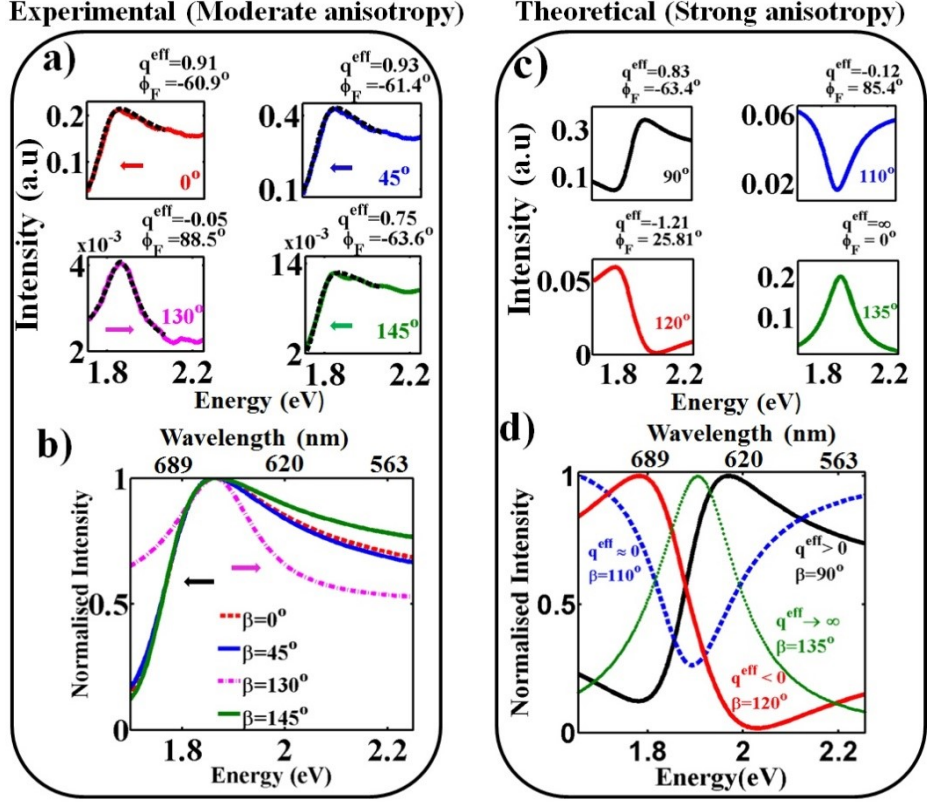

**Figure 4:** Tuning of Fano spectral asymmetry in **(a, b)** moderately anisotropic experimental waveguided plasmonic crystal sample, and **(c, d)** theoretical predictions for a strongly anisotropic system. The pre-selected state is elliptical ($\alpha = +45^0, \Phi_\alpha - \Phi_\beta \sim 0.8\ rad$, $\mathbf{S_\alpha} = [1\ 0\ 0.697\ -0.717]^T$) in **(a, b)** and linear ($\alpha = +45^0, \Phi_\alpha - \Phi_\beta = 0$, $\mathbf{S_\alpha} = [1\ 0\ 1\ 0]^T$) in **(c, d)**. Post-selections are at different linear polarization angles β ($\mathbf{S_\beta} = [1\ \cos2\beta\ \sin2\beta\ 0]^T$). The spectral variation of the scattered intensities are shown in (a) and (c) and the changes in the line shapes are highlighted in (b) and (d). The fitted (using Eq. 3) parameters ($q^{\text{eff}}$, $\varphi_F$) are noted. In (a, b), the pre and post-selections are done on experimental $M_{\text{Pol}}$ matrix. In (c, d), these are simulated using Eq. (4) and (6) with ($q_x = 1.5, q_y = 0.5, B_x = B_y = B = 0.6$) and ($E_{0,x} = 1.893\ eV, \lambda_{0,x} = 655$ nm; $E_{0,y} = 1.882\ eV, \lambda_{0,y} = 659$ nm).